\documentclass[12pt]{book}

\usepackage[dvips]{graphicx,color}
\usepackage{makeidx,hori,zon}
\def\be{\begin{equation}}
\def\ee{\end{equation}}
\def\bea{\begin{eqnarray}}
\def\eea{\end{eqnarray}}

\makeauthorindex

\BookTitle{Inflating Horizon of Particle Astrophysics and Cosmology}
\CopyRight{\copyright 2005 by Universal Academy Press, Inc.\ and Yamada Science Foundation}

\begin{document}

\pagenumbering{arabic}

\chapter{
Challenges for String Gas Cosmology}

\author{%
Robert H. BRANDENBERGER\\
{\it Department of Physics, McGill University,
3600 rue Universit\'e, Montreal, QC, H3A 2T8,
Canada\\
rhb@hep.physics.mcgill.ca}}
%
%
\AuthorContents{R. Brandenberger} 

\AuthorIndex{Brandenberger}{R.} 

\section*{Abstract}

In spite of the phenomenological successes of the inflationary universe
scenario, the current realizations of inflation making use of scalar
fields lead to serious conceptual problems which are reviewed in this
lecture. String theory may provide an avenue towards addressing these
problems. One particular approach to combining string theory and
cosmology is String Gas Cosmology. The basic principles of this
approach are summarized.

\section{Introduction}

The current paradigm of early universe cosmology is the inflationary
universe scenario \cite{Guth,Sato} (see also \cite{Starob1,Brout} for
earlier ideas). Most implementations of this scenario are based
on the existence of a slowly rolling scalar field whose energy-momentum
tensor is dominated by the contribution of the field potential energy 
which drives a period of accelerated (and in fact in most models
almost exponential) expansion. In spite of the impressive phenomenological
successes of this paradigm in predicting the spectrum of density
perturbations and the angular power spectrum of cosmic microwave
background (CMB) anisotropies, the scalar field-driven inflationary
universe scenario suffers from some important conceptual problems
\cite{RHBrev3,Review1}.
Addressing these problems is one of the goals of superstring
cosmology.

One of the conceptual problems of scalar field-driven inflation
is the {\it amplitude problem}, namely the
fact that in the simplest realizations of the model, a hierarchy
of scales needs to be present in order to be able to obtain the
observed small amplitude of the primordial anisotropies (see
e.g. \cite{Adams} for a fairly general analysis of this problem).

A more serious problem is the {\it trans-Planckian problem} \cite{RHBrev3}.
As can be seen from the space-time diagram of Figure 1, 
provided that the period of inflation lasted sufficiently
long (for GUT scale inflation the number is about 70 e-foldings),
then all scales inside of the Hubble radius today started out with a
physical wavelength smaller than the Planck scale at the beginning of
inflation. Now, the current theory of cosmological perturbations
(the theory used to calculate the spectra of density fluctuations
and microwave anisotropies) is based
on Einstein's theory of General Relativity coupled to a simple
semi-classical description of matter. It is clear that these
building blocks of the theory are inapplicable on scales comparable
and smaller than the Planck scale. Thus, the key
successful prediction of inflation (the theory of the origin of
fluctuations) is based on an incomplete analysis since we know that 
new physics must enter
into a correct computation of the spectrum of cosmological perturbations.
The key question is as to whether the predictions obtained using
the current theory are sensitive to the specifics of the unknown
theory which takes over on small scales. Toy model calculations
using modified dispersion relations \cite{Jerome1,Niemeyer} have
shown that the predictions are in fact sensitive to Planck-scale
physics, thus opening up the exciting possibility to test Planck-scale
and string-scale physics in current observations (see \cite{Jerome2}
for a review with references to other approaches towards exploring
this ``trans-Planckian window of opportunity''.
\begin{figure}
\begin{center}
\includegraphics[height=6cm]{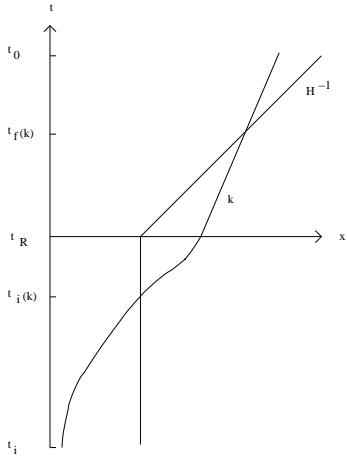}
\caption{Space-time diagram (sketch) showing the evolution
of scales in inflationary cosmology. The vertical axis is
time, and the period of inflation lasts between $t_i$ and
$t_R$, and is followed by the radiation-dominated phase
of standard big bang cosmology. During exponential inflation,
the Hubble radius $H^{-1}$ is constant in physical spatial coordinates
(the horizontal axis), whereas it increases linearly in time
after $t_R$. The physical length corresponding to a fixed
comoving length scale labelled by its wavenumber $k$ increases
exponentially during inflation but increases less fast than
the Hubble radius (namely as $t^{1/2}$), after inflation.}
\end{center}
\label{fig:1}       
\end{figure}

A third problem is the {\it singularity problem}. It was known for a long
time that standard Big Bang cosmology cannot be the complete story of
the early universe because of
the initial singularity, a singularity which is unavoidable when basing
cosmology on Einstein's field equations in the presence of a matter
source obeying the weak energy conditions (see e.g. \cite{HE} for
a textbook discussion). Recently, the singularity theorems have been
generalized to apply to Einstein gravity coupled to scalar field
matter, i.e. to scalar field-driven inflationary cosmology \cite{Borde}.
It is shown that in this context, a past singularity at some point
in space is unavoidable. Thus we know, from the outset, that scalar
field-driven inflation cannot be the ultimate theory of the very
early universe.

The Achilles heel of scalar field-driven inflationary cosmology is,
however, the {\it cosmological constant problem}. We know from
observations that the large quantum vacuum energy of field theories
does not gravitate today. However, to obtain a period of inflation
one is using the part of the energy-momentum tensor of the scalar field
which looks like the vacuum energy. In the absence of a convincing
solution of the cosmological constant problem it is unclear whether
scalar field-driven inflation is robust, i.e. whether the
mechanism which renders the quantum vacuum energy gravitationally
inert today will not also prevent the vacuum energy from
gravitating during the period of slow-rolling of the inflaton 
field. Note that the
approach to addressing the cosmological constant problem making use
of the gravitational back-reaction of long range fluctuations
(see \cite{RHBrev4} for a summary of this approach) does not prevent
a long period of inflation in the early universe.

Finally, a key challenge for inflationary cosmology is to find a
well-motivated candidate for the scalar field which drives inflation,
the inflaton. Ever since the failure of the model of {\it old inflation}
\cite{Guth,Sato}, it is clear that physics beyond the Standard Model
of particle physics must be invoked. 

It is likely that string theory will provide ideas which allow us
to successfully address some of the abovementioned problems of
the current versions of inflationary cosmology. Foremost, since one
of its goals is to resolve space-time singularities,
string theory has a good chance of providing a nonsingular cosmology (and {\it
string gas cosmology}, the scenario explored below, indeed has the
potential of resolving the cosmological singularity). Since string
theory should describe physics on all scales, it should be possible
to compute the spectrum of cosmological 
perturbations in a consistent way within string theory, thus opening the {\it
trans-Planckian window of opportunity}. Finally, string theory contains
many light scalar fields, good candidates for inflation.

In this lecture, I present a review of {\it String Gas Cosmology} (SGC),
an attempt to apply ideas of superstring theory to construct a new
paradigm of early universe cosmology. The main idea of SGC is to apply
new symmetries (T-duality) and new degrees of freedom (string winding
modes) which are central to string theory and to apply them to early
universe cosmology. In the following section, I will present the basics
of SGC. Next, the equations which describe the dynamics of SGC will
be discussed. I will end with one section describing recent progress
in the context of SGC towards stabilizing all of the moduli which
describe the volume and shape of the extra spatial dimensions which
critical superstring theory predicts (this topic will
be reviewed in more detail in a accompanying write-up \cite{Review3}), 
and with a section listing some outstanding challenges.
 
\section{Challenges for String Cosmology}

An immediate problem which arises when trying to connect string theory
with cosmology is the {\bf dimensionality problem}. Superstring theory
is perturbatively consistent only in ten space-time dimensions, but we
only see three large spatial dimensions. The original approach to 
addressing this problem is to assume that the six extra dimensions are
compactified on a very small space which cannot be probed with our
available energies. However, from the point of view of cosmology,
it is quite unsatisfactory not to be able to understand why it is
precisely three dimensions which are not compactified and why the compact
dimensions are stable. Brane world cosmology \cite{branereview} provides
another approach to this problem: it assumes that we live on a
three-dimensional brane embedded in a large nine-dimensional space.
Once again, a cosmologically satisfactory theory should explain
why it is likely that we will end up exactly on a three-dimensional
brane (for some interesting work addressing this issue see
\cite{Mahbub,Mairi,Lisa}).

Finding a natural solution to the dimensionality problem is thus one
of the key challenges for superstring cosmology. This challenge has
various aspects. First, there must be a mechanism which singles out
three dimensions as the number of spatial dimensions we live in.
Second, the moduli fields which describe the volume and the shape of
the unobserved dimensions must be stabilized (any strong time-dependence
of these fields would lead to serious phenomenological constraints).
This is the {\bf moduli problem} for superstring cosmology. As
mentioned above, solving the {\bf singularity problem} is another of
the main challenges. These are the three problems which {\bf string
gas cosmology} \cite{BV,TV,ABE} explicitly addresses at the present
level of development.

In order to make successful connection with late time cosmology,
any approach to string cosmology must also solve the 
{\bf flatness problem}, namely make sure that the three large
spatial dimensions obtain a sufficiently high entropy (size) to
explain the current universe. Finally, it must provide a
mechanism to produce a nearly scale-invariant spectrum of
nearly adiabatic cosmological perturbations.

Since superstring theory leads to many light scalar fields, it is
 possible that superstring cosmology will provide a
convincing realization of inflation (see e.g. \cite{stringinflation}
for reviews of recent work attempting to obtain inflation in the
context of string theory). However, it is also possible that
superstring cosmology will provide an alternative to cosmological
inflation, maybe along the lines of the Pre-Big-Bang \cite{PBB}
or Ekpyrotic \cite{KOST} scenarios. The greatest challenge for
these alternatives is to solve the flatness problem (see e.g.
\cite{Pyro}). 

\section{Heuristics of String Gas Cosmology}

In the absence of a non-perturbative formulation of string theory,
the approach to string cosmology which we have suggested,
{\it string gas cosmology} \cite{BV,TV,ABE} (see also \cite{Perlt}),
is to focus on symmetries and degrees of freedom which are new to
string theory (compared to point particle theories) and which will
be part of a non-perturbative string theory, and to use
them to develop a new cosmology. The symmetry we make use of is
{\bf T-duality}, and the new degrees of freedom are 
{\bf string winding modes}.

We take all spatial directions to be toroidal, with
$R$ denoting the radius of the torus. Strings have three types
of states: {\it momentum modes} which represent the center
of mass motion of the string, {\it oscillatory modes} which
represent the fluctuations of the strings, and {\it winding
modes} counting the number of times a string wraps the torus.
Both oscillatory and winding states are special to strings as
opposed to point particles. 

The energy of an oscillatory mode is independent of $R$, momentum
mode energies are quantized in units of $1/R$, i.e.
\be
E_n \, = \, n {1 \over R} \, ,
\ee
and winding mode energies are quantized in units of $R$:
\be
E_m \, = \, m R \, ,
\ee
where both $n$ and $m$ are integers.

The T-duality symmetry is a symmetry of the spectrum of string
states under the change
\be \label{Tdual}
R \, \rightarrow \, {1 \over R}
\ee
in the radius of the torus (in units of the string length $l_s$).
Under such a change, the energy spectrum of string states is
invariant: together with the transformation (\ref{Tdual}), winding
and momentum quantum numbers need to be interchanged
\be \label{Tdual2}
(n, m) \, \rightarrow \, (m, n) \, .
\ee
The string vertex operators are consistent with this symmetry, and
thus T-duality is a symmetry of perturbative string theory. Postulating
that T-duality extends to non-perturbative string theory leads
\cite{Pol} to the need of adding D-branes to the list of fundamental
objects in string theory. With this addition, T-duality is expected
to be a symmetry of non-perturbative string theory.
Specifically, T-duality will take a spectrum of stable Type IIA branes
and map it into a corresponding spectrum of stable Type IIB branes
with identical masses \cite{Boehm}.

We choose
the background to be dilaton gravity. It is crucial to include the
dilaton in the Lagrangian, firstly since
the dilaton arises in string perturbation theory at the same level
as the graviton (when calculating to leading order in the string
coupling and in $\alpha'$), and secondly because it is only the action of
dilaton gravity (rather than the action of Einstein gravity)
which is consistent with the T-duality symmetry. Given this
background, we consider an ideal gas of matter made up of all
fundamental states of string theory, in particular including
string winding modes.
 
Any physical theory requires initial conditions. We assume that
the universe starts out small and hot. For simplicity, we take
space to be toroidal, with radii in all spatial directions given by
the string scale. We assume that the initial energy density 
is very high, with an effective temperature which is close
to the Hagedorn temperature, the maximal temperature
of perturbative string theory. 

Based on the T-duality
symmetry, it was argued \cite{BV} that the cosmology resulting from SGC
will be non-singular. For example, as the background radius $R$
varies, the physical temperature $T$ will obey the symmetry
\be
T(R) \, = \, T(1/R)
\ee
and thus remain non-singular even if $R$ decreases to zero.
Similarly, the length $L$ measured by a physical observer will
be consistent with the symmetry (\ref{Tdual}), hence realizing
the idea of a minimal physical length \cite{BV}.

Next, it was argued \cite{BV} that in order for spatial sections to become
large, the winding modes need to decay. This decay, at least
on a background with stable one cycles such as a torus, is only
possible if two winding modes meet and annihilate. Since string
world sheets have measure zero probability for intersecting in more
than four space-time dimensions, winding modes can annihilate only
in three spatial dimensions (see, however, the recent
caveats to this conclusion based on the work of \cite{Kabat3}). 
Thus, only three spatial dimensions
can become large, hence explaining the observed dimensionality of
space-time. As was shown later \cite{ABE}, adding branes to
the system does not change these conclusions since at later
times the strings dominate the cosmological dynamics.
Note that in the three dimensions which are becoming large there
is a natural mechanism of isotropization as long as some winding
modes persist \cite{Watson1}.

\section{Equations for String Gas Cosmology}

The equations of SGC are based on coupling an ideal gas of all
string and brane modes, described by an energy density $\rho$
and pressures $p_i$ in the i'th spatial direction, 
to the background space-time of
dilaton gravity. They follow from varying the action
\be
S \, = \, {1 \over {2 \kappa^2}} \int d^{10}x \sqrt{-g} e^{-2 \phi}
\bigl[{\hat R} + 4 \partial^{\mu} \phi \partial_{\mu} \phi \bigr] + S_m \, ,
\ee
where $g$ is the determinant of the metric, ${\hat R}$ is the Ricci scalar,
$\phi$ is the dilaton,
$\kappa$ is the reduced gravitational constant in ten dimensions, 
and $S_m$ denotes the matter action. The
metric appearing in the above action is the metric in the
string frame. In the case of a homogeneous
and isotropic background given by
\be
ds^2 \, = \, dt^2 - a(t)^2 d{\bf x}^2 \, ,
\ee
the three resulting equations (the
generalization of the two Friedmann equations plus the equation
for the dilaton) in the string frame are
\cite{TV} (see also \cite{Ven})
\bea
-d {\dot \lambda}^2 + {\dot \varphi}^2 \, &=& \, e^{\varphi} E 
\label{E1} \\
{\ddot \lambda} - {\dot \varphi} {\dot \lambda} \, &=& \,
{1 \over 2} e^{\varphi} P \label{E2} \\
{\ddot \varphi} - d {\dot \lambda}^2 \, &=& \, {1 \over 2} e^{\varphi} E \, ,
\label{E3}
\eea
where $E$ and $P$ denote the total energy and pressure, respectively,
$d$ is the number of spatial dimensions, and we have introduced the
logarithm of the scale factor 
\be
\lambda(t) \, = \, {\rm log} (a(t))
\ee
and the rescaled dilaton
\be
\varphi \, = \, 2 \phi - d \lambda \, .
\ee

Note that the contribution to the pressure from winding modes
is negative, wherease that from the momentum modes is positive.
Thus, from the second of the above equations it follows immediately
that a gas of strings containing both stable winding and
momentum modes will lead to the stabilization of the
radius of the torus: windings prevent expansion, momenta
prevent the contraction. The right hand side of the equation
can be interpreted as resulting from a confining potential for
the scale factor. Note that this behavior is a consequence
of having used dilaton gravity rather than Einstein gravity
as the background. The dilaton is evolving at the time when
the radius of the torus is at the minimum of its potential.

The above background equations thus demonstrate that, in order
for any spatial dimensions to be able to grow large, the
winding modes circling this dimension must be able to
annihilate. In the case of three spatial dimensions, the
interaction of string winding modes can be described in
analogy with the interaction of cosmic strings. Two winding strings
with opposite orientations
can intersect, producing closed loops with vanishing winding
as a final state. The equations which describe the energy
transfer between winding and non-winding strings are given
in analogy to the case of cosmic strings 
(see e.g. \cite{VilShell,HK,RHBrev5} for reviews). First, we
split the energy density in strings into the density in winding
strings
\be
\rho_w(t) \, = \, \nu(t) \mu t^{-2} \, ,
\ee
where $\mu$ is the string mass per unit length, and $\nu(t)$ is
the number of strings per Hubble volume, and into the density in
string loops
\be
\rho_l(t) \, = \, g(t) e^{-3 (\lambda(t) - \lambda(t_0))} \, ,
\ee
where $g(t)$ denotes the comoving number density of loops, normalized
at a reference time $t_0$. In terms of these variables, the
equations describing the loop production from the interaction of
two winding strings are \cite{BEK}
\bea
{{d \nu} \over {dt}} \, &=& \, 
2 \nu \bigl( t^{-1} - H \bigr) - c' \nu^2 t^{-1} \label{E4} \\
{{dg} \over {dt}} \, &=& \, c' \mu t^{-3} \nu^2 
e^{3 \bigl( \lambda(t) - \lambda(t_0) \bigr)} \label{E5}
\eea
where $c'$ is a constant, which is of order unity for cosmic strings
but which depends on the dilaton in the case of fundamental strings
\cite{Kabat3}.

The system of equations (\ref{E1}, \ref{E2}, \ref{E3}, \ref{E4}, \ref{E5})
was studied in \cite{BEK} (see also \cite{Campos03}). It was
verified that the presence of a large initial density of string winding
modes causes any initial expansion of $a(t)$ to come to a halt. Thereafter,
$a(t)$ will decrease. The resulting increase in the density of winding
strings will lead to rapid loop production, and the number of winding
strings will decrease, then allowing $a(t)$ to start expanding again.
In \cite{BEK}, this initial evolution of $a(t)$ was called ``loitering''.
In \cite{BEK}, the analysis was performed using a constant value of $c'$.
Taking into account the dilaton dependence of $c'$, one finds
\cite{Kabat3} that the annihilation mechanism and resulting liberation
of the three large dimensions only works if the initial value of the
dilaton is sufficiently large.

\section{Progress in String Gas Cosmology}

A key issue in all approaches to string cosmology is the question of
{\bf moduli stabilization}. The challenge is to fix the shape and
volume moduli of the compact dimensions and to fix the value of
the dilaton. Moduli stabilization is essential to obtain a 
consistent late time cosmology.

There has recently been a lot of progress on the issue of moduli
stabilization in SGC, progress which will be reviewed in
detail in \cite{Review3}. In a first study \cite{Watson2}, the stabilization
of the radii of the extra dimensions (the ``radion'' degrees of freedom)
was studied in the string frame. It was shown that, as long as there are
an equal number of string momentum and winding modes about the
compact directions, the radii are dynamically stabilized at the 
self-dual radius. The dilaton, however, is in general evolving in time.

For late time cosmology, it is crucial to show that the radion degrees
of freedom are stabilized in the Einstein frame. Obstacles towards
achieving this goal were put forward in \cite{BattWat,Aaron,Tirtho,Damien}.
However, if the spectrum of string states contains modes which are
massless at the self-dual radius (which is the case for Heterotic
but not for Type II string theory), then these modes generate an
effective potential for the radion which has a minimum with vanishing
energy at the self-dual radius and thus yields radion stabilization
\cite{Subodh1,Subodh2} (see also \cite{Watson3}). 
As shown in these references, the
radion stabilization mechanism is consistent with late time
cosmology (e.g. fifth force constraints).
The same massless modes also yield  stabilization
of the shape moduli \cite{Edna,Sugumi}. The outstanding challenge in this
approach is to stabilize the dilaton (for some ideas see \cite{Subodh3}). 

There has been other important work on SGC. For example, in \cite{Easther1}
is was shown that topologically stable one cycles (such as exist for
a toroidal background) are not necessary for the success of SGC in
predicting that only three spatial dimensions can become large. The
results generalize to certain orbifold backgrounds for which the
winding strings are long-lived but not absolutely stable. On the other
hand, for other corners of the hypothetical M-theory moduli space, for
example 11-d supergravity, the resulting cosmology may not single out
three as the number of large spatial dimensions since there are no 
fundamental one-dimensional objects \cite{Easther2} (see, however,
the ideas in \cite{Stephon}). The brane thermodynamics during the
early stages of SGC was studied in \cite{Borunda}. The effects of
wrapped branes were studied in \cite{Kaya}, the effects of background
fluxes were studied in \cite{Campos}, and generalizations to 
Calabi-Yau backgrounds and backgrounds
with non-vanishing spatial curvature were considered in \cite{Damien2,Biswas}.
For other work on moduli stabilization in SGC see \cite{Kaya2}.

An important concern is the danger that string windings lead to new
instabilities towards fluctuations in the size of the radion as a
function of the spatial coordinates of the large three-dimensional
space. In \cite{Watson4,Watson5} it was shown that - at least before
dilaton stabilization - no such instabilities arise.

Finally, recently there has been interest in the possibility
that the massive modes of SGC could contribute to the dark
matter \cite{Gubser,BattWat} and/or dark energy \cite{Tirtho3}.

\section{Problems for String Gas Cosmology}

As mentioned in the previous section, an important outstanding
problem in SGC is to find a mechanism to fix both the dilaton
and the radion at the same time. This problem is presumably
a reflection in SGC of the difficulties faced in other approaches
to string cosmology, for example in the Type IIB flux compactification
models \cite{GKP} when attempting to fix all moduli 
(in those models, it is the
volume modulus which is difficult to stabilize).

Since SGC assumes as initial conditions that all spatial sections
are of string scale, it appears that inflation of the three large
dimensions is required in order to solve the entropy problem
\cite{Guth}. Obtaining inflation from string gas ideas has proved
to be difficult (see \cite{Moshe,Easson2,Tirtho2} for some recent ideas), 
and a major
challenge is to obtain inflation consistent with moduli stabilization.
In any cosmology which proposes an alternative to inflation (and it
is possible that SGC will lead to such an alternative) the most
difficult problem is, once again, to solve the entropy problem.

It would also be nice to put SGC on a firmer footing. The current
considerations are rather heuristic - e.g. they rely
on an artificial splitting between a classical background
space-time and stringy matter, a splitting which is bound to
become a bad approximation on stringy distance scales, 
and it would be nice to base
an improved scenario on a more consistent analysis of strings
in non-trivial background space-times.

\centerline{\bf Acknowledgements}

I would like to thank the organizers of the 59th Yamada Conference
(and in particular Profs. Y. Suto and J. Yokoyama) for inviting me to speak
and for their wonderful hospitality during my visit in Tokyo. I
also wish to thank Prof. S. Mukohyama for his help, and
Thorsten Battefeld for comments on this draft. This research is
supported by an NSERC Discovery Grant and by the Canada Research
Chairs program.



\end{document}